\documentclass[twocolumn]{aastex62}
\usepackage{natbib}
\usepackage{rotating}
\newcommand{\msun}{$M_{\odot}$}

\newcommand{\editone}[1]{#1}

\newcommand{\refs}[1]{\color{cyan}{[refs]}\color{black}}

\graphicspath{{./}{figures/}}

\received{}
\revised{}
\accepted{}
\submitjournal{ApJ}

\shorttitle{Sulfur in planet-forming rocks}
\shortauthors{Kama et al.}

\begin{document}

\title{Abundant refractory sulfur in protoplanetary disks}

\correspondingauthor{Mihkel Kama}
\email{mkama@ast.cam.ac.uk}

\author[0000-0003-0065-7267]{Mihkel Kama}
\affil{Institute of Astronomy, University of Cambridge, Madingley Road, Cambridge CB3 0HA, UK}

\author{Oliver Shorttle}
\affil{Institute of Astronomy, University of Cambridge, Madingley Road, Cambridge CB3 0HA, UK}

\author{Adam S. Jermyn}
\affil{Institute of Astronomy, University of Cambridge, Madingley Road, Cambridge CB3 0HA, UK}
\affil{Kavli Institute for Theoretical Physics, University of California at Santa Barbara, Santa Barbara, CA 93106, USA}

\author{Colin P. Folsom}
\affil{IRAP, Universit\'e de Toulouse, CNRS, UPS, CNES, 31400, Toulouse, France}

\author{Kenji Furuya}
\affil{Center for Computational Sciences, University of Tsukuba, 1-1-1 Tennoudai, 305-8577, Tsukuba, Japan}

\author{Edwin A. Bergin}
\affil{Department of Astronomy, University of Michigan, 1085 S. University Avenue, Ann Arbor, MI 48109, USA}

\author{Catherine Walsh}
\affil{School of Physics and Astronomy, University of Leeds, Leeds, LS2 9JT, UK}

\author{Lindsay Keller}
\affil{ARES, Code XI3, NASA/JSC, Houston, TX 77058, USA}



\begin{abstract}

Sulfur is one of the most abundant elements in the Universe, with important roles in astro-, geo-, and biochemistry.  Its main reservoirs in planet-forming disks have previously eluded detection: gaseous molecules only account for $<1\,$\% of total elemental sulfur, with the rest likely in either ices or refractory minerals.  Mechanisms such as giant planets can filter out dust from gas accreting onto disk-hosting stars.  For stars above 1.4 solar masses, this leaves a chemical signature on the stellar photosphere that can be used to determine the fraction of each element that is locked in dust.  Here, we present an application of this method to sulfur, zinc, and sodium.  We analyse the accretion-contaminated photospheres of a sample of young stars and find $(89\pm8)\,$\% of elemental sulfur is in refractory form in their disks. The main carrier is much more refractory than water ice, consistent with sulfide minerals such as FeS.

\end{abstract}



\section{Introduction}\label{sec:intro}

Sulfur is an atomic gas in the interstellar medium (ISM), but is found entirely in rocks in the inner solar system. Its main reservoir in protoplanetary disks has thus far eluded observation. We study a sample of $16$ young, disk-hosting stars, listed in Appendix~\ref{apx:data}.  Trapping of large dust grains in their disks, evidenced by radial gaps and cavities in the distribution of millimetre-wavelength continuum emission \citep{Andrewsetal2009, vanderMareletal2016, Andrewsetal2018}, has been shown to correlate with a depletion of refractory elements like iron on the stellar surface \citep{Kamaetal2015}. In this paper, we measure the refractory fraction of sulfur, zinc, and sodium by comparing their behaviour with the overall level of dust depletion in the accreting inner disk material.

\begin{figure*}[!ht]
\centering
\includegraphics[clip=,width=1.0\linewidth]{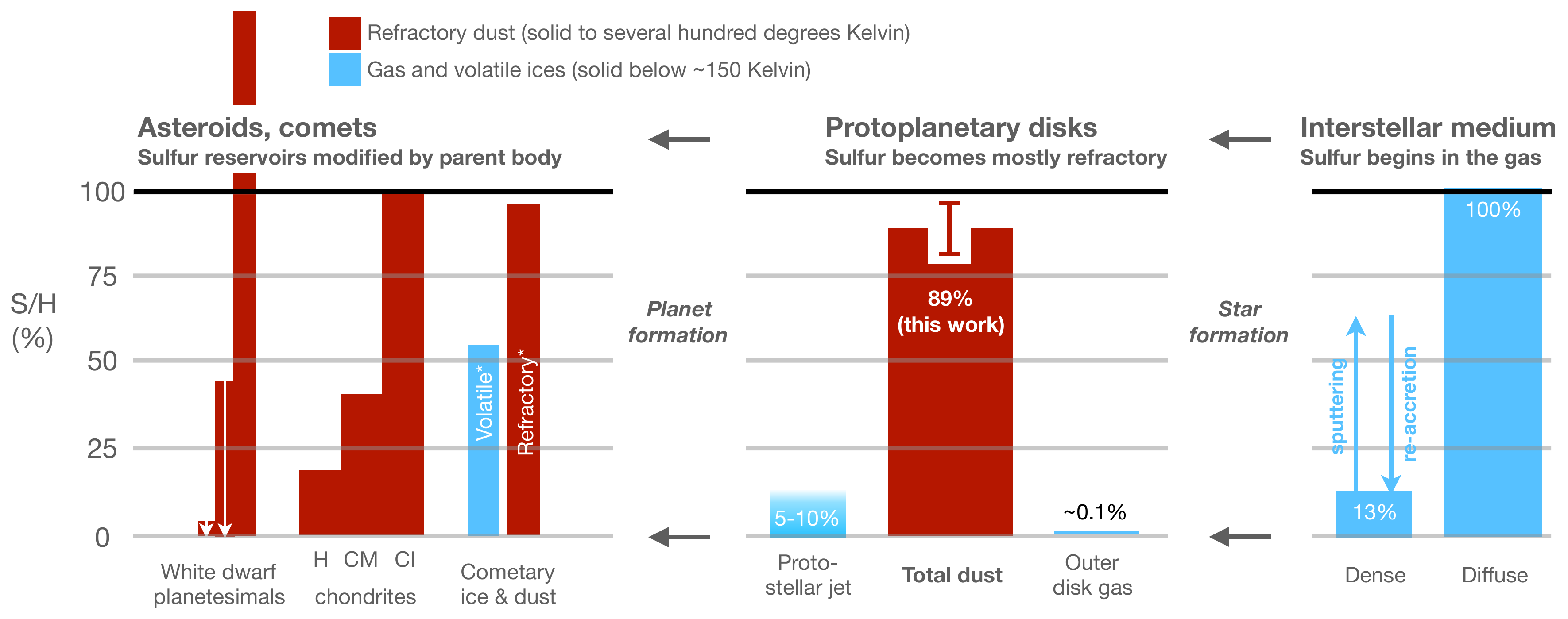}
\caption{\emph{Fractional importance of volatile (blue) and refractory (red) reservoirs of sulfur, as measured in environments sampling different stages of the star and planet formation process \citep{Gaensickeetal2012, Xuetal2013, Xuetal2017, WassonKallemeyn1988, Calmonteetal2016, Andersonetal2013, Dutreyetal1997, Dutreyetal2011, Wakelametal2004b, Fuenteetal2010, MartinDomenechetal2016, Jenkins2009}. The normalizations of the panels are, from right to left: the highest, super-solar gas-phase abundance in the diffuse ISM \citep{Jenkins2009}; the reference sulfur abundance in early-type stars as determined from young open clusters \citep{Fossatietal2011,Martinetal2017}; and solar S/H, to which we scale via silicon for dust and oxygen for ices \citep{Asplundetal2009}, assuming refractories have a solar abundance of Si and cometary ices have solar O. White arrows covering solid bars indicate upper limits.}}
\label{fig:abunbar}
\end{figure*}

The cosmic journey of sulfur is summarized in Figure~\ref{fig:abunbar}, where we trace sulfur from the interstellar medium, through disks (this work), and finally into rocky and icy planetesimals.  Sulfur is chiefly synthesized in \editone{Type~II supernovae \citep[e.g.][]{WoosleyWeaver1995, RydeLambert2005} and supernova remnants contain gas-phase molecules such as SO \citep{Matsuuraetal2017}.  In (post-) asymptotic giant branch stars, }observations reveal a high abundance of gas-phase molecular carriers such as CS, SO, SiS, and H$_{2}$S, but also sulfide mineral grains such as FeS and MgS \citep[e.g.][]{Honyetal2002, Danilovichetal2016, Danilovichetal2017, Danilovichetal2018}. Refractory sulfur is converted to gas by the enhanced sputtering rate of solid sulfide by ions, a process which is much less efficient for metals and silicates \citep[][]{Kelleretal2013}. Once it has entered the diffuse interstellar medium (ISM), sulfur is entirely in the gas phase \citep{Josephetal1986, Jenkins2009}. 

As sulfur moves into denser regions of the ISM, its gas-phase abundance drops to $\sim 13\,$\% of the diffuse ISM value \citep{Jenkins2009}.  Little is known about the reservoirs consuming sulfur along the path from the ISM to the denser star-forming molecular cloud cores and protoplanetary disks, in which the vast majority of it has eluded observation.  Our measurement fills this gap.  What was known until now is that volatile ices such as H$_{2}$S and OCS account for $\lesssim 4\,$\% of the total S inventory in star-forming cores \citep[][Fig.~\ref{fig:abunbar}]{Geballeetal1985, Smith1991, Palumboetal1995, Boogertetal1997, Boogertetal2015}, while in shocked regions in jets from embedded protostars only $5$ to $10\,$\% of total sulfur is in the gas phase \citep{Andersonetal2013}.  In planet-forming disks, gas-phase H$_{2}$S, CS, and SO only account for $\lesssim 1\,$\% of all sulfur \citep{Dutreyetal1997, Wakelametal2004b, Fuenteetal2010, Dutreyetal2011, MartinDomenechetal2016, Semenovetal2018}, while tentatively detected spectral features of FeS have suggested a poorly-quantifiable refractory component \citep{Kelleretal2002, Lisseetal2007}.

\section{Analysis}

Our analysis relies on dust trapping in disks, which removes some refractory material material before can accrete onto the central star; and on the fact that early-type stars are dominated by diffusive rather than convective mixing in their envelopes, which allows recently accreted disk material to dominate the composition of their surface.  These considerations allow us to extract new information from composition data of disk-hosting stars, gathered from the literature and our own previous work.

Stars more massive than $1.4\,$M$_{\odot}$ have radiative envelopes where mixing is dominated by slow diffusion, as opposed to the faster convective mixing which occurs in lower-mass stars.  Their observable photosphere mass, of order $10^{-10}\,$M$_{\odot}$, can be entirely replaced on a timescale of days at disk accretion rates $\dot{M}_{\rm acc}\sim 10^{-8}$ to $10^{-7}\,$M$_{\odot}\,$yr$^{-1}$ \citep{JermynKama2018}.  Measuring the stellar surface composition is thus a tool for studying the elemental composition of material recently accreted from a circumstellar disk \citep[the Contaminated A-stars Method, \texttt{CAM};][]{Kamaetal2015, JermynKama2018}.  Surface layers built up on the star from dust-depleted material in the protoplanetary disk stage form a small fraction of the total mass of the star, and are lost through diffusive and rotational mixing with the bulk of the star within a million years after the disk dissipates \citep{JermynKama2018}.  


\begin{figure*}[!ht]
\centering
\includegraphics[clip=,width=1.0\linewidth]{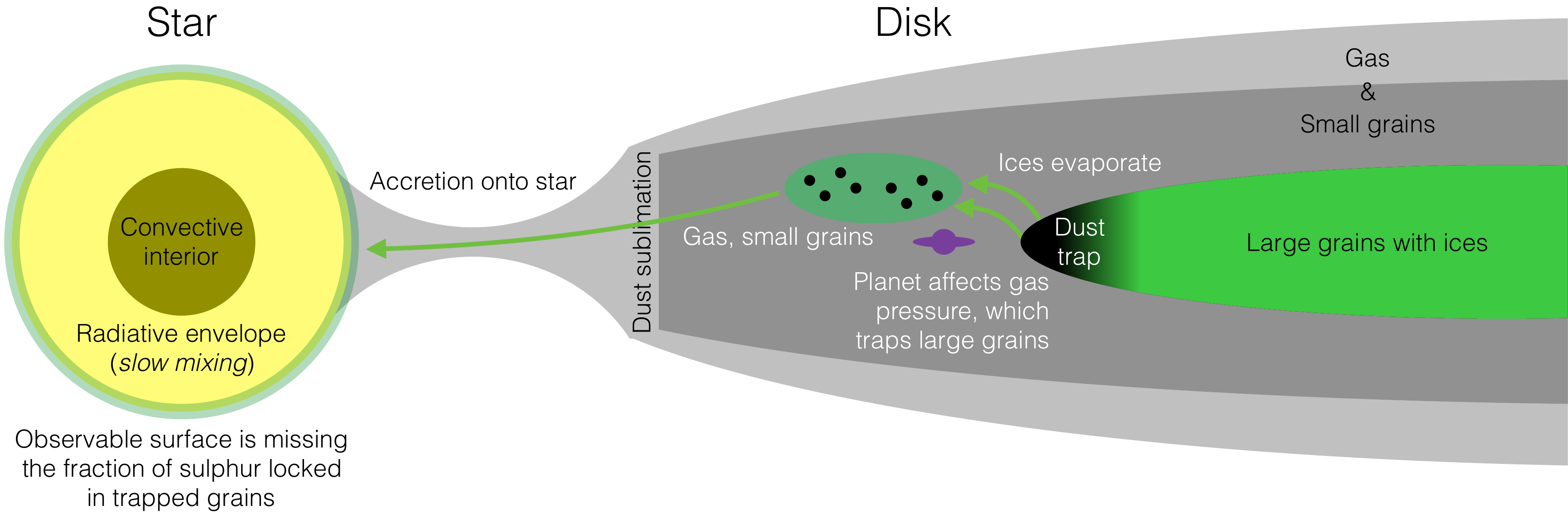}
\caption{\emph{Accretion of gas, ice, and dust grains from a disk onto a young star.  While gas and smaller dust grains flow freely towards the star, a radial gas pressure bump, potentially induced by a planet, filters away the larger dust grains. This prevents some dust from moving inwards and accreting, depriving material reaching the star of refractory elements.  The chemical signature is visible in the photosphere of stars more massive than $1.4\,$\msun, which lack a convective envelope, as long as the dust trap exists and material from that disk region is reaching the star.}}
\label{fig:diskstar}
\end{figure*}

\subsection{Stellar surface and inner disk composition}

We gathered surface elemental composition data for a sample of $16$ young, disk-hosting stars of stellar spectral type B9 through F4 from the literature \citep[][see Appendix~\ref{apx:data} for details]{Folsometal2012, Kamaetal2016b}.  We assume each star and disk starts out with the same reference composition (X/H)$_{\rm ref}$, where X is an element, for which we adopted the mean abundances from studies of nearby, young open clusters by \citet[][NGC\,6250, age $26\,$Myr]{Martinetal2017} and \citet[][NGC\,5460, age $158\,$Myr]{Fossatietal2011}\footnote{Even though the open cluster early-type stars themselves have measurement scatter of a few tenths of dex, the disk-hosting stars which have dust gaps or cavities have surface abundances of refractory elements extending well below this range.}.  \editone{All elemental abundance trends we report and analyse below are present in the disk-hosting star sample alone, but the robustness is increased by using a reference set of disk-free stars.  This also gives us enough data to discard disk hosts with poorly quantified errorbars.}

A parcel of material in a disk starts out with a fraction $f_{\rm X}$ of each element X in the refractory (dust grains) component and the rest in volatiles (gas, ice).  For hydrogen $f_{\rm H}=0$ and for strong refractories such as iron and titanium we fix $f_{\rm Fe,Ti}=1$, expecting them to be entirely refractory and essentially measuring the overall level of dust depletion in the inner disk.  For each star, we follow \citet{JermynKama2018} and Section~\ref{apx:model} in calculating the mixing fraction $f_{\rm ph}\in{[0,1]}$ of recently accreted material.  The results below are not substantially different if we assume the photosphere is totally replaced by recent accretion in all the stars ($f_{\rm ph}=1)$.  Each star-disk system also has a dust depletion fitting parameter, $\delta_{\rm dust}$, which scales the refractory component of each element to account for dust retained in the disk.  The abundance of X in disk material that makes it onto the star is (X/H)$_{\rm disk}$.  We perform statistical inference to obtain $f_{\rm X}$ for O, S, Zn, and Na using Equations~\ref{eq:xhstar} and \ref{eq:acctot}. Appendix~\ref{apx:data} describes the disk-hosting and reference stars, and Section~\ref{apx:model} the inference model relating the measured composition of accreting stars to the inner disk material.

 \editone{In order to, firstly, test our results for robustness and, secondly, to consider zinc for which all open cluster stars failed the quality criteria, we also ran our analysis using the solar composition from \citet{Asplundetal2009}. We note that the solar composition is not an optimal reference for the early-type star abundances, which were determined using a different set of spectral features and stellar atmospheric models, and may be further modulated by Galactic chemical evolution.  We have made no attempt in this work to quantify these confounding effects, focussing in our main analysis on the direct comparison of open cluster and disk-hosting early-type stars.}

\subsection{Dust trapping as a chemical isolation tool}

Our method is insensitive to the precise mechanism of dust trapping in disks, but segregation of some dust from a parcel of disk material generally requires a radially non-monotonous midplane gas pressure.  Trapping is evidenced by resolved observations which show millimetre-sized grains adjacent to dust-poor regions \citep[e.g.][]{Andrewsetal2009, vanderMareletal2016, Andrewsetal2018}.  Such dust cavities and gaps in disks correlate with a lowered abundance of refractory elements accreted onto the central star \citep{Kamaetal2015} and planet-induced dust traps match the observations \citep{Pinillaetal2012}.  A planet embedded in a protoplanetary disk causes a pressure bump to form radially outward of the planet (Figure~\ref{fig:diskstar});  gas drag and centrifugal forces then conspire to accumulate larger dust particles in the local pressure maximum \citep{Birnstieletal2016, Pinillaetal2012}.  As large grains dominate the dust mass, trapping has a major effect on the total elemental composition of the disk material that reaches the star \citep{Kamaetal2015}.  Building on the relation between stellar photospheric compositions and dust disk structure, we can open a new window onto the reservoirs of each element in the zone of terrestrial planet formation.

\begin{figure*}[!ht]
\centering
\includegraphics[clip=,width=0.33\linewidth]{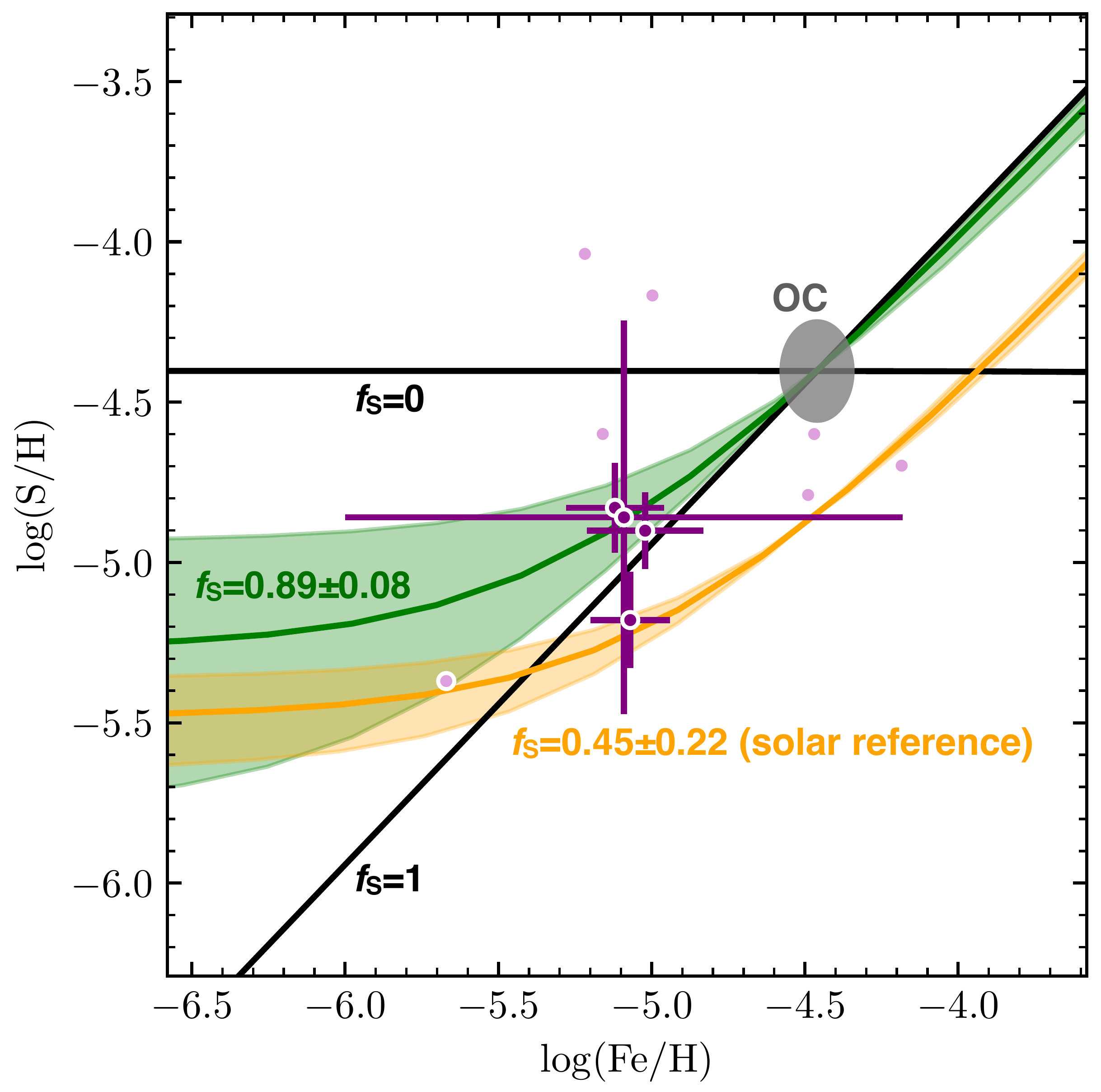}
\includegraphics[clip=,width=0.66\linewidth]{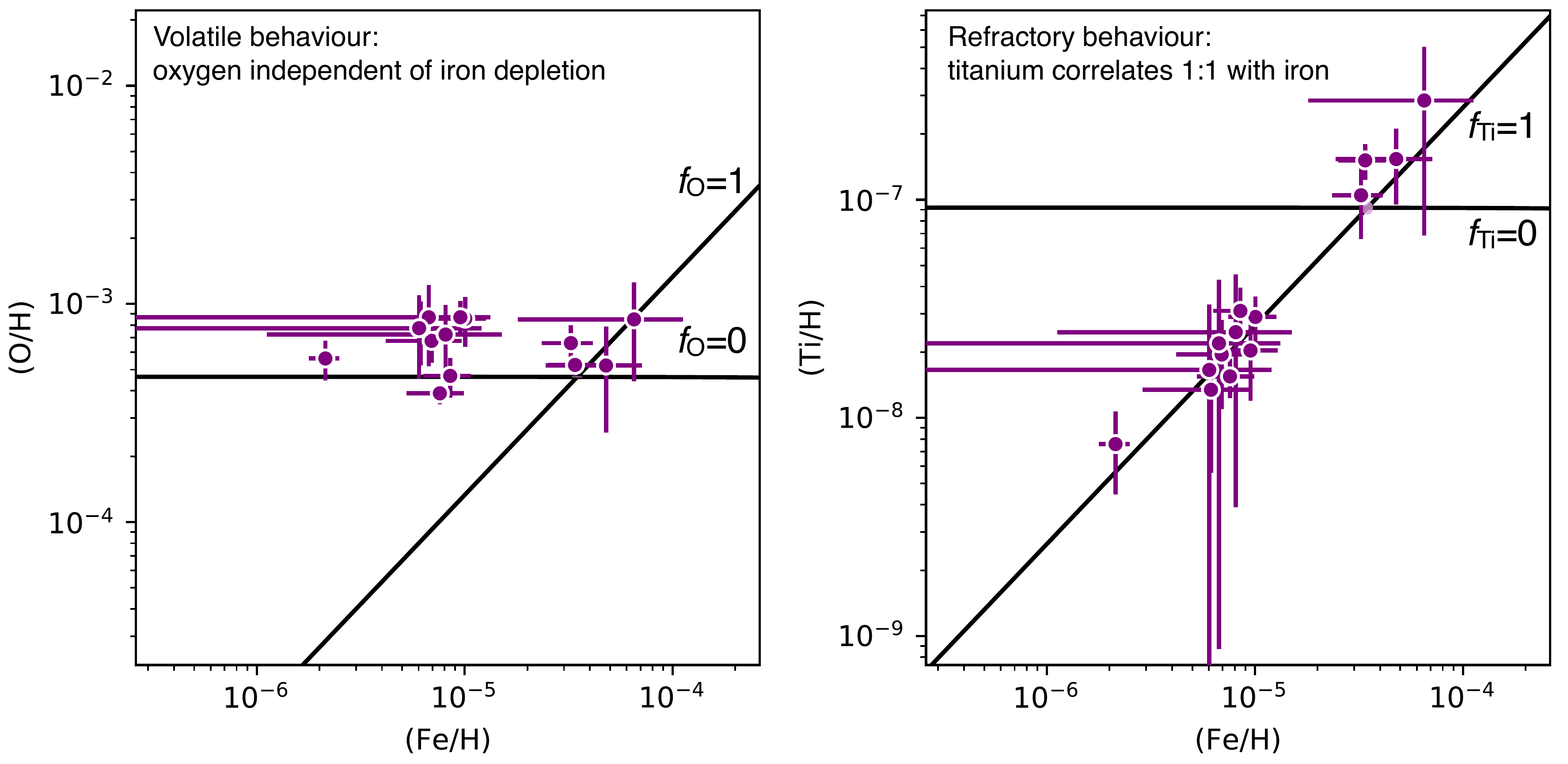}
\caption{Abundance of sulfur (left-hand panel), oxygen (middle), and titanium (right-hand) versus iron, normalised to hydrogen.  For disk-hosting stars, the inferred composition of inner disk material is shown (dark purple).  Stars with poorly quantified uncertainties were excluded (light purple, no errorbars shown). The reference composition derived from open cluster data is defined to lie at the intersection of the $100\,$\% volatile and refractory behaviours ($f_{\rm X}=0$ and $1$, solid black lines, ``OC''). Our adopted fit uses the open cluster reference (green, $1\sigma$ shaded). A solar-reference fit (orange) is shown for comparison.  See Appendix~\ref{apx:data} for stellar composition references.}
\label{fig:fehsh}
\end{figure*}

\subsection{Volatile, refractory, and intermediate elements}

The observed surface compositions of our compiled sample of disk-hosting stars and the open cluster reference are given in Table~\ref{tab:sources}. We use the analysis described in Section~\ref{apx:model} to infer the elemental composition of the inner disk for the accreting, young stars.  These inferred abundances are shown alongside the open cluster reference in Figure~\ref{fig:fehsh}, to illustrate the volatile behaviour of oxygen, the refractory behaviour of titanium and iron, and the behaviour of sulfur.  We show correlation plots of iron \citep[$50\,$\% condensation temperature $T_{\rm c}=1334\,$K,][]{Lodders2003} with sulfur ($664\,$K), oxygen ($180\,$K), and titanium ($1582\,$K).  Titanium correlates 1:1 with iron, consistent with both elements being entirely in refractory form. Removing a fraction of the dust from an accreting mass parcel lowers Fe and Ti in the same proportion. While $\approx30\,$\% of disk-hosting A-type stars have a low surface abundance of these refractory elements \citep{Folsometal2012}, among all A-type stars the fraction is only $2\,$\% \citep{GrayCorbally1998, Paunzen2001}, confirming that this is a short-lived surface contamination effect from disk accretion \citep{Kamaetal2015}. Sulfur shows more scatter than titanium, but nonetheless displays a strong correlation with iron. This allows us to infer the fraction of sulfur locked in dust particles. 

\subsection{Refractory fraction of an element}\label{apx:model}

We define (X/H)$_{\rm ref}$ as the reference abundance of element X in the material from which stars form, i.e. their bulk composition.  This initial composition is mostly built up prior to the protoplanetary disk stage in which our disk hosts are and we assume it to be the same for the accreting, disk-hosting stars (sample stars) and nearby, young open cluster stars (reference stars) of the same effective temperature range (see also Appendix~\ref{apx:data}).  We determined the reference composition by taking an unweighted average of those open cluster stars for which a given element had been measured.  Uncertainties on the \editone{open cluster} reference abundances were calculated as the sample standard deviation, and these determine the size of the reference composition uncertainty ellipses in Figure~\ref{fig:fehsh}.

We next define $f_{\rm X}$ as the fraction of X locked in refractories and $\delta_{\rm d}$ as the level of change of the refactory dust mass in the accretion stream, for example due to a planet-induced dust cavity, where $\delta_{\rm d}<1$ describes a depletion of dust and $\delta_{\rm d}>1$ describes a dust excess.  We assume the accretion stream composition directly samples the total elemental composition of the inner disk from which the stream originates.  We then get the following for the abundance of X accreted from the disk onto the stellar photosphere:
\begin{equation}
\left( \frac{\rm X}{\rm H} \right)_{\rm disk} = \left[(1-f_{\rm X}) + f_{\rm X}\, \delta_{\rm d}\right] \times \left( \frac{\rm X}{\rm H} \right)_{\rm ref}.
\label{eq:xhstar}
\end{equation}

Now let $f_{\rm ph}$ be the mass fraction of accreted material in the photosphere of a star~\citep{JermynKama2018}. Then the observed stellar composition is related to the composition of the accreting material as
\begin{equation}
\left( \frac{\rm X}{\rm H} \right)_{\star} = f_{\rm ph}\left( \frac{\rm X}{\rm H} \right)_{\rm disk} + \left(1-f_{\rm ph}\right) \left( \frac{\rm X}{\rm H} \right)_{\rm ref}.
\label{eq:acctot}
\end{equation}
We determined $f_{\rm ph}$ using the methods in~\citet{JermynKama2018} for those stars with a measured accretion rate, rotational velocity $v \sin i$, and surface temperature.  Equation~\ref{eq:acctot} was then used to solve for the composition of the inner disk.  Assuming $\delta_{\rm d}$, \editone{defined above as a dust mass scaling factor,} affects the refractory component of all elements equally \editone{and thus cancels out}, the accreted abundance of elements X and Y is related through
\begin{equation}
\left( \frac{\rm Y}{\rm H} \right)_{\rm disk} = \left( \frac{\rm Y}{\rm H} \right)_{\rm ref}\times \left[ 1 + \frac{f_{\rm Y}}{f_{\rm X}}\, \left( \frac{\left( \frac{\rm X}{\rm H} \right)_{\rm disk}}{\left( \frac{\rm X}{\rm H} \right)_{\rm ref}} - 1 \right) \right]
\label{eq:yxrel}
\end{equation}


In our main analysis, we use the Bayesian Multinest sampling algorithm~\citep{doi:10.1111/j.1365-2966.2007.12353.x,doi:10.1111/j.1365-2966.2009.14548.x,1306.2144} to obtain posterior probability distributions on the free parameters, which are $f_{\rm S}$ and $f_{\rm Zn}$ globally, as well as $\delta_{\rm d}$ and $f_{\rm ph}$ for each star. \editone{Zinc} is excluded here due to a lack of good data as described in Appendix~\ref{apx:data}, but see below for a tentative result on $f_{\rm Zn}$. We take uniform priors over $f_{\rm S}$ and $f_{\rm Na}$ from $0$ to $1$. We take a log-uniform prior for $\delta_{\rm d}$ from $10^{-3}$ to $10^3$. Finally, we take the prior distribution of $f_{\rm ph}$ to be log-normal centered on the mean calculated computed using the methods in~\citet{JermynKama2018} and with variance determined by propagating uncertainties in the inputs to those methods.
The resulting Gaussian is cut off at three standard deviations in each direction or at the logical boundary $f_{\rm ph} \leq 1$ of the domain, whichever is more restrictive. After sampling we marginalise over uncertainties in the accreted fraction $f_{\rm ph}$ and in the observed abundances.

\section{Results}\label{sec:results}

We fit for the refractory fraction of O (which behaves essentially as a volatile), S, Zn, and Na.  Elements with $T_{\rm c}>1000\,$K are assumed to be entirely locked in dust and were assigned $f_{\rm X}=1$. This includes elements such as Fe, Mg, Si, and Ti.  All uncertainties quoted below, and elsewhere in the paper, are given at $1\,\sigma$.

We obtain $f_{\rm S}=(89\pm8)\,$\% for sulfur and $f_{\rm Na}=(35\pm16)\,$\% for sodium.  This is the first measurement of the fraction of these elements locked in refractory reservoirs in protoplanetary disks.
Note that we imposed no prior on the relation between $f_{\rm S}$ and $f_{\rm Na}$. Equilibrium condensation calculations suggest $f_{\rm S}<f_{\rm Na}$, and imposing this constraint as a prior to check our results, we find $f_{\rm S}=89\,$\% and $f_{\rm Na}=97\,$\%. This illustrates the robustness of our result for sulfur, and underlines the large uncertainty in $f_{\rm Na}$.

We repeated this analysis using solar abundances from \citet{Asplundetal2009} as the stellar reference.  \editone{Keeping in mind an unknown systematic offset from the early-type star abundances, this} enabled us to also consider $f_{\rm Zn}$, for which we took the prior to be uniform from $0$ to $1$.
This yields $f_{\rm S} = (45\pm 22)\,$\%, $f_{\rm Zn} = (52\pm 34)\,$\% and $f_{\rm Na} = (77 \pm 18)\,$\%.
The strong dependence of our results on the reference abundance points to the need to choose this from as similar a population of stars as possible.
The increased uncertainty in these estimates relative to those from the field star reference reflects the generally worse fit obtained using the solar reference.
This may also be seen in the Bayesian evidence, which was $\log L \approx -1401$ for the field star reference and $\log L \approx -9912$ for the solar reference.
The two figures are not perfectly comparable because the solar reference enabled more elements to be used in the fit, which generally lowers $\log L$ by a factor of order the ratio of the number of observations used.
This disagreement is much larger than that effect, though, because the solar abundance is not as reflective of the underlying bulk abundances of the open cluster A-type and young Herbig~Ae/Be stars.

We carried out another check by performing orthogonal distance regression fitting of the sulfur-iron correlation, using solar abundances as the reference point. This yields a refractory sulfur fraction $f_{\rm S}=(75\pm8)\,$\%, consistent with the Multinest results.  Due to the self-consistent posteriors on the reference composition, we consider Multinest fitting with free reference values the superior approach and have highlighted the resulting $f_{\rm S}$ value ($89\pm8)\,$\%) as the most reliable.

The summarised statistics from our analysis are available upon request  as supplementary data in the Multinest JSON format.  The analysis methods were verified on test data, as discussed in Appendix~\ref{apx:testing}.


\section{Discussion}

\subsection{Nature of the sulfur reservoirs}
The high O abundance in all our accretion-contaminated stars points towards complete ice evaporation in, or prior to, the dust depletion location in the disks. At the H$_{2}$O snowline, any S-bearing volatile ices (H$_{2}$S, OCS, SO, SO$_{2}$) would also have evaporated, enabling their gas-phase transport onto the central star. If there is significant refractory S however, its trapping in the pressure bump would mean that material accreted to the star is S-poor, which is what we find.

This refractory sulfur must have a sublimation temperature higher than that of H$_{2}$O \citep[T$_{\rm sub}\approx100\,$K,][]{Collingsetal2004}.  This constraint is not met by the volatile species H$_{2}$S, CS, SO, and SO$_{2}$ which would all co-desorb with water and accrete onto the star; but it is consistent with sulfide minerals, such as FeS \citep[T$_{\rm sub}\approx655\,$K --][]{Lodders2003, Larimer1967a} and with sulfur chains S$_{n}$, where $n=2,\ldots,8$. Sulfur chains are more volatile than sulfide minerals and experimentally less volatile than water \citep[][]{JimenezEscobarMunozCaro2011}. According to these works, S$_{2}$ desorbs at $\gtrsim 150\,$K and S$_{3}$ at $\gtrsim 260\,$K. In contrast to efficient sulfide formation from gaseous H$_{2}$S and solid Fe \citep{Laurettaetal1996}, however, the abundance of stable sulfur chains only reaches $\lesssim 1\,$\% of all S nuclei in chemical models \citep{Charnley1997, DruardWakelam2012} and $\sim 6\,$\% in experiments \citep{JimenezEscobarMunozCaro2011, Woodsetal2015}, adding to the weight of evidence favouring sulfide minerals as the main reservoir.

While cometary ices contain S$_{n}$ at the percent level \citep{Calmonteetal2016} and cometary dust is high in FeS content \citep{Jessbergeretal1988, Westphaletal2009}, meteoritic rocks which sample inner solar system planetesimals show a mix of sulfides and chains: linear and cyclical sulfur chains intermixed with carbon have a comparable abundance to that of sulfide minerals \citep{OrthousDaunayetal2010}. The chain and sulfide groups together account for most of the meteoritic sulfur, although the relative abundances of various groups varies, perhaps due to a combination of different intrinsic abundances and parent body alteration history. We therefore favour the interpretation that the refractory S component in dust prior to incorporation in planetesimals is predominantly FeS and other sulfide minerals.

\subsection{An excess of volatile oxygen}

Regardless of the level of depletion of the refractory elements Fe and Ti in disk-hosting stars, the O abundance does not vary.  We find $f_{\rm O}=(2\pm2)\,$\%. This is low, as condensation models put $\sim 23\,$\% of O atoms in silicates \citep{Lodders2003}, and warrants further study.  Most oxygen occurs in volatile form in a solar-composition mixture \citep{Lodders2003}.  These considerations suggest that O is dominated by volatile carrier molecules and that the dust traps preventing refractories from accreting are always warm enough that the most abundant O-bearing ices -- CO, CO$_{2}$, and H$_{2}$O -- evaporate. The water snowline can be far out in the disks of these luminous stars.  Detailed models of the HD~100546 disk show that H$_{2}$O ice can evaporate at the outer edge of its dust-depleted cavity, at $\sim15\,$au \citep{Kamaetal2016b}.  So a potential explanation to the apparent non-dependence of oxygen on the level of dust depletion would be if higher levels of dust depletion correlated positively with an increased delivery of water molecules through the dust trapping region.

\subsection{Longevity of dust traps and stellar abundance anomalies}
All of the disks around refractory-poor stars in our sample have dust-depleted regions in their inner disk on $\lesssim 10\,$au scales \citep{Kamaetal2015}.  The viscous spreading timescale\footnote{The viscous evolution timescale is $t_{\nu}=(\alpha\,\Omega_{\rm K})^{-1}\times (h/r)^{-2}$, where $\alpha$ is the Shakura-Sunyaev viscosity parameter, $\Omega_{\rm K}$ the local Kepler time, and $(h/r)$ the scaleheight-to-radius ratio.} from $10\,$au is $t\sim10^{5}\,$yr, assuming the viscosity parameter is $\alpha=10^{-2}$.  Thus, the closer a dust-depleted zone is to the star, the more likely it is that the material currently observed to accrete onto the star has a physical memory of that particular zone.  The stellar surface composition will change again on the viscous timescale if disk evolution alters the amount of dust reaching the star; or over $\sim10^{6}\,$years after accretion stops, as diffusive mixing with deeper layers of the star dilutes away the chemical fingerprint of the surface layers \citep{TurcotteCharbonneau1993, JermynKama2018}.

\subsection{Implications for exoplanets and the solar system}

Volatile and refractory sulfur reservoirs are observable in the products of planet formation: in solar system asteroids and comets, and in disrupted planetesimals around post-main-sequence stars (Fig.~\ref{fig:abunbar}, left-hand panel).  The meteorites thought to most closely resemble primordial material from the inner ten astronomical units of our protoplanetary disk are the CI chondrites \citep{WassonKallemeyn1988}.  The S/Si ratio in these is very nearly equal to that in the sun \citep{Asplundetal2009}.  Meteorites such as CM or H chondrites that are fragments of processed, variably melted, differentiated parent bodies can have a substantially lower sulfur content.  A similar, but larger range of sulfur abundance is inferred for exo-planetesimals accreting onto white dwarfs \citep{Gaensickeetal2012, Xuetal2013, Xuetal2017}, although the phase of sulfur in these bodies is uncertain and there may be measurement systematics as well as true initial abundance effects which push the highest inferred white dwarf planetesimal S/H ratio to a super-solar value (Fig.~\ref{fig:abunbar}).

The \emph{Rosetta} spacecraft measured an elemental S/O ratio consistent with solar in both the ice \emph{and} the dust of comet 67P/Churyumov-Gerasimenko \citep[][Fig.~\ref{fig:abunbar}]{Calmonteetal2016}. While cometary ices sample the volatile species of the outer disk, a significant fraction of the refractory cometary dust is thought to be reprocessed particles from the inner disk that were radially mixed outwards, based on their sulfide mineral content and the presence of crystalline silicates \citep[e.g.][]{Westphaletal2009}. The high abundance of sulfur-bearing cometary ices suggests that sulfur was predominantly volatile at the onset of proto-solar disk formation, while its high abundance in rocky planetesimals supports the suggestion that sulfide minerals were produced in high abundance from H$_{2}$S or isomers of OCS reacting with solid Fe as the intensively accretion-heated inner few astronomical units of the proto-solar nebula cooled below $700\,$K \citep{Kerridge1976, HaugenSterten1971, StertenHaugen1973, Laurettaetal1996, Chambers2009}. Our measurement strongly supports such models, wherein the reprocessing of volatile sulfur species in disks to refractory minerals is a general process.

Differentiation of rocky planets such as Earth and Mercury is partly controlled by their sulfur content \citep{Malavergneetal2014, Laurenzetal2016}, and the element plays a major role in atmospheric chemistry and may have helped shield early life from ultraviolet radiation \citep{HapkeNelson1975, WinickStewart1980, Zhangetal2010, Kastingetal1989}. Our result provides a new constraint for protostellar and -planetary chemical models, and for observations tallying sulfur reservoirs in disks with ALMA and other sub-millimetre telescopes \citep{Dutreyetal1997, Wakelametal2004b, Fuenteetal2010, Dutreyetal2011, MartinDomenechetal2016, BoothAetal2018}. We predict $\leq(11\pm8)\,$\% of total sulfur in the gas and ice phase in the rocky planet formation zone of protoplanetary disks, which is predominantly further out than the $T_{\rm dust}\sim700\,$K destruction limit of FeS. If a large fraction of the refractory sulfur is in chains, the gas-phase sulfur abundance will be elevated above $11\,$\% at radii inwards of $\approx 200\,$K before peaking inside of $700\,$K.

In the near future, sulfur in the form of hydrogen sulfide (H$_{2}$S) may be observationally constrained in Hot Jupiter atmospheres with the \emph{James Webb} Space Telescope and other facilities. In the case of core accretion with a pure gas envelope, JWST may find very little H$_{2}$S in the atmospheres of Hot Jupiters, unless late accretion of planetesimals has provided significant contamination.

\section{Conclusions}

We have identified $(89\pm8)\,$\% of all sulfur in refractory form in the innermost regions of protoplanetary disks.  The main reservoir must be substantially more refractory than H$_{2}$O ice, which strongly favours sulfide minerals (e.g. FeS) over sulfur chain molecules (S$_{n}$). Almost all elemental sulfur is thus available for direct incorporation in rocky planetesimals in the inner few to ten astronomical units around stars of $\approx2$ to $3\,$\msun. The rest is in volatile ices or gas-phase species.

This measurement was made possible by the use of accretion contamination on the surfaces of early-type stars as a probe of circumstellar material \citep{JermynKama2018}. The radiative envelopes of such stars prevent recently accreted material from rapidly mixing with the deeper layers of the envelope, so the photospheric composition can easily be dominated by fresh material at accretion rates typical for protoplanetary disks.

\acknowledgments
MK gratefully acknowledges funding from the European Union's Horizon 2020 research and innovation programme under the Marie Sklodowska-Curie Fellowship grant agreement No 753799. ASJ is funded by the Gordon and Betty Moore Foundation through Grant GBMF7392 and the National Science Foundation under Grant No. NSF PHY-1748958. CW acknowledges financial support from the University of Leeds and the Science and Technology Facilities Council (STFC; grant number ST/R000549/1).

\bibliography{sulfur}

\appendix

\section{Data}\label{apx:data}

\subsection{Stars and disks}
Stellar elemental abundances for our analysis were taken from the studies of \citet[][Herbig~Ae/Be stars]{Folsometal2012}, \citet[][the Herbig~Ae/Be star HD~100546]{Kamaetal2016b}, \citet[][NGC~5460]{Fossatietal2011}, and \citet[][NGC~6250]{Martinetal2017}. These studies used similar quality data and a similar methodology, whereby the stellar photospheric properties were determined simultaneously and self-consistently with the elemental abundances. 

The Multinest fitting (Section~\ref{apx:model}) to obtain the refractory fractions $f_{\rm O}$, $f_{\rm S}$, $f_{\rm Zn}$, and $f_{\rm Na}$ was carried out globally over all stars and all elements, with a few exclusions. For each element-element combination, we excluded all stars for which a relevant abundance was determined from a single spectral feature and had an errorbar assigned from the researchers' experience as opposed to from spectral model fitting. We also entirely excluded open cluster stars with an effective temperature over $200\,$K away from the limits of the disk-hosting star sample. The excluded stars for sulfur (plotted as light-coloured symbols without errorbars in Figure~\ref{fig:fehsh}) were HD~31648, HD~36112, HD~68695, HD~179218, and HD 244604 in \citet{Folsometal2012}; and HD~123269, UCAC~11105213, and UCAC~11105379 in \citet{Fossatietal2011}. Temperature excluded UCAC~12284506 in \citet{Martinetal2017}. Mostly, the abundances of the excluded stars are $3\,\sigma$-consistent with our final best fit. Finally, a sample study of abundances in Herbig~Ae/Be stars by \citet{AckeWaelkens2004} which did include sulfur was excluded from consideration due to their different abundance-fitting methodology. These authors used stellar T$_{\rm eff}$, R$_{\star}$ and $\log{(g)}$ values from the literature and fitted only for the elemental abundances. A comparison of abundances for the stars in common between this study and that of \citet{Folsometal2012} gives confidence that they are mostly within $3\sigma$ of each other.

The baseline abundance for each element was calculated as the mean of all open cluster stars which passed our exclusion criteria for that element, as described in the previous paragraph. We found solar abundances to be an unsatisfactory baseline for a number of elements, most likely due to a combination of Galactic chemical evolution in the $\sim4\,$Gyr separating the young sun from the birth of the relevant open cluster stars; and the different spectroscopic data and stellar models used to measure solar and early-type field star abundances.

\begin{sidewaystable}[!ht]
\centering
\begin{tabular}{ l c c c c c c c }
\hline\hline
Star		& $\rm \log_{10}(\frac{O}{H})$ & $\rm \log_{10}(\frac{S}{H})$	&  $\rm \log_{10}(\frac{Zn}{H})$	& $\rm \log_{10}(\frac{Na}{H})$	& $\rm \log_{10}(\frac{Fe}{H})$	& $\rm \log_{10}(\frac{Ti}{H})$ & $\log_{10}(f_{\rm ph})$	\\
\hline
HD 31648	& $-3.24\pm0.05^{*}$	& $-4.456\pm0.30^{*}$	& --	& $-5.68\pm0.25^{*}$ 	& $-4.43\pm0.13$ & $-6.78\pm0.09$ 	& $-6.680\times10^{-3}$ \\
HD 36112	& $-3.14\pm0.10^{*}$	& $-4.75\pm0.16^{*}$ & $-7.76\pm0.40^{*}$ 	& $-5.54\pm0.15^{*}$ & $-4.45\pm0.14$ 	& $-6.94\pm0.20$ & $-6.771\times10^{-3}$ \\
HD 68695	& $-3.13\pm0.10$ & $-4.56\pm0.30^{*}$ 	& -- & $-6.16\pm0.40^{*}$ 	& $-5.12\pm0.22$ & $-7.67\pm0.25$ & $-6.784\times10^{-3}$ \\
HD 100546	& $-3.25\pm0.10$ & $-5.37\pm0.50^{*}$ & -- & -- & $-5.67\pm0.08$ & $-8.12\pm0.23$ & $-6.904\times10^{-3}$ \\
HD 101412	& $-3.08\pm0.09$ & $-4.92\pm0.12$ & -- & $-5.52\pm0.15^{*}$ & $-5.04\pm0.19$ & $-7.71\pm0.23$ & $-6.713\times10^{-3}$ \\
HD 139614	& $-3.29\pm0.10^{*}$ & $-5.14\pm0.15$ & $-8.26\pm0.30^{*}$ & $-6.10\pm0.12$ & $-5.03\pm0.13$ & $-7.47\pm0.14$ & $-6.696\times10^{-3}$ \\
HD 141569	& $-3.01\pm0.10$ & -- & -- & $\leq-5.16$ & $-5.21\pm0.32$  & $-7.70\pm0.32$ & $-4.141\times10^{-1}$ \\
HD 142666	&  $-3.14\pm0.15^{*}$ & $-4.66\pm0.15$ & $-7.86\pm0.30^{*}$ & $-5.73\pm0.15^{*}$ & $-4.80\pm0.11$ & $-7.35\pm0.20$ & $-3.071\times10^{-1}$ \\
HD 144432	& $-3.13\pm0.10^{*}$ & $-4.78\pm0.05$ & $-7.53\pm0.20$ &  $-5.82\pm0.09$ 	& $-4.66\pm0.09$ & $-7.22\pm0.16$ & $-5.114\times10^{-1}$ \\
HD 163296	& $-3.27\pm0.15$ & -- & -- & $-5.56\pm0.50^{*}$ & $-4.35\pm0.15$ & $-6.88\pm0.10$ & $-4.178\times10^{-1}$ \\
HD 169142	& $-3.34\pm0.13$ & $-5.06\pm0.12$ & $-8.67\pm0.08^{*}$ & $-6.14\pm0.08^{*}$ & $-5.09\pm0.11$ & $-7.61\pm0.10$ & $-3.525\times10^{-1}$ \\
HD 179218	& $-3.06\pm0.13$ & $-4.16\pm0.40^{*}$ & -- & $-5.46\pm0.30^{*}$ & $-4.99\pm0.13$ & $-7.53\pm0.12$ & $-6.804\times10^{-3}$ \\
HD 244604	& $-3.19\pm0.08$ & $-4.44\pm0.15^{*}$ & -- & $-5.36\pm0.30$ & $-4.31\pm0.24$ & $-6.77\pm0.31$ & $-7.291\times10^{-1}$ \\
HD 245185	& $-3.13\pm0.17$ & -- & -- & -- & $-5.23\pm0.33$ & $-7.89\pm0.38$ & $-6.848\times10^{-3}$ \\
HD 278937	& $-3.37\pm0.05$ & $-4.79\pm0.14$ & $\leq7.96$ & $-6.26\pm0.30^{*}$ & $-5.08\pm0.16$ & $-7.77\pm0.10$ & $-6.928\times10^{-3}$ \\
T Ori	 	& $-3.12\pm0.12$ & $-4.06\pm0.30^{*}$ & -- & $-6.06\pm0.30^{*}$ & $-4.94\pm0.10$ & $-7.51\pm0.17$ & $-4.437\times10^{-1}$ \\
\hline
Reference (ref)	& $-3.33\pm0.07$ & $-4.33\pm0.28$ & $\ldots$ & \editone{$-5.86$} & $-4.65\pm0.47$ & $-7.10\pm0.62$ & \\
\hline
Solar ($\odot$) & $-3.31\pm0.05$ & $-4.88\pm0.03$ & $-7.44\pm0.05$ & $-5.76\pm0.04$ & $-4.50\pm0.04$ & $-7.05\pm0.05$ & \\
\hline
\end{tabular}
\caption{Measured photospheric abundances of O, S, Zn, Na, Fe, and Ti from \citet{Folsometal2012} for our sample stars except HD~100546, which is from \citet{Kamaetal2016b}. The reference abundances are the mean of open cluster stars \editone{(with errors which are actually the sample standard deviation)}, except for zinc which is referenced to the \citet{Asplundetal2009} solar value ($^{\odot}$). The photospheric contamination fraction was calculated for each star following \citet{JermynKama2018}. $^{*}$ marks abundances determined from a single spectral feature or a very limited spectral range -- these values are excluded from our fitting but shown in all plots as lighter-coloured symbols.}
\label{tab:sources}
\end{sidewaystable}


\subsection{Chondrites}
We adopted the chondrite meteorite abundances from \citet{WassonKallemeyn1988}. The most relevant here are the CI chondrites, which are thought to originate in undifferentiated, minimally processed parent bodies.

\subsection{Comets}
We adopt the S/O ratios measured for the volatile (evaporated ice) component of comet 67P/Churyumov-Gerasimenko by the ESA \emph{Rosetta} mission, as reported in \citet{Calmonteetal2016}. No other comet has yet been studied in comparable detail. The species included in the volatile S/O ratio calculation are H$_{2}$S, S, OCS, S$_{2}$, SO$_{2}$, SO, H$_{2}$O, CO, CO$_{2}$, and O$_{2}$. For the abundance of sulfur in the refractory dust of comet 1P/Halley, we use measurements from the PUMA mass-spectrometer \citep{Jessbergeretal1988} on Vega-1 \citep{Sagdeevetal1986}; and for comet 81P/Wild~2, X-ray spectroscopic measurements of a dust grain \citep{Westphaletal2009} returned by the \emph{Stardust} mission \citep{Brownleeetal2006}.



\section{Testing}\label{apx:testing}
To verify our inference methods we generated an artificial dataset composed of stars with depleted elemental abundances relative to solar abundances according to Equation~\ref{eq:acctot}.  For these purposes different elements were assigned different refractory fractions and different stars were assigned different dust depletion factors.  These abundances and the baseline were then contaminated with log-normal noise of various amplitudes.  For each star a random subset of elements was dropped from the dataset to reflect the fact that in actual observations not all elements have an abundance determination.

We performed our inference analysis on this artificial dataset and were able to successfully infer the dust depletion factors as well as refractory fractions to within a tolerance of the same magnitude as the noise used to contaminate the sample.

\end{document}